\input harvmac
\noblackbox

\def\tilde{\widetilde}
\newcount\figno
\figno=0
\def\fig#1#2#3{
\par\begingroup\parindent=0pt\leftskip=1cm\rightskip=1cm\parindent=0pt
\baselineskip=11pt
\global\advance\figno by 1
\midinsert
\epsfxsize=#3
\centerline{\epsfbox{#2}}
\vskip 12pt
\centerline{{\bf Figure \the\figno:} #1}\par
\endinsert\endgroup\par}
\def\figlabel#1{\xdef#1{\the\figno}}

\def\np#1#2#3{Nucl. Phys. {\bf B#1} (#2) #3}
\def\pl#1#2#3{Phys. Lett. {\bf B#1} (#2) #3}

\def\cqg#1#2#3{Class. Quantum Grav. {\bf #1} (#2) #3}


\font\cmss=cmss10
\font\cmsss=cmss10 at 7pt
\def\rlx{\relax\leavevmode}
\def\inbar{\vrule height1.5ex width.4pt depth0pt}
\def\IC{\relax\,\hbox{$\inbar\kern-.3em{\rm C}$}}
\def\IN{\relax{\rm I\kern-.18em N}}
\def\IP{\relax{\rm I\kern-.18em P}}
\def\ZZ{\rlx\leavevmode\ifmmode\mathchoice{\hbox{\cmss Z\kern-.4em Z}}
 {\hbox{\cmss Z\kern-.4em Z}}{\lower.9pt\hbox{\cmsss Z\kern-.36em Z}}
 {\lower1.2pt\hbox{\cmsss Z\kern-.36em Z}}\else{\cmss Z\kern-.4em
 Z}\fi} 
\def\IZ{\relax\ifmmode\mathchoice
{\hbox{\cmss Z\kern-.4em Z}}{\hbox{\cmss Z\kern-.4em Z}}
{\lower.9pt\hbox{\cmsss Z\kern-.4em Z}}
{\lower1.2pt\hbox{\cmsss Z\kern-.4em Z}}\else{\cmss Z\kern-.4em
Z}\fi}

\def\narrowplus{\kern -.04truein + \kern -.03truein}
\def\narrowminus{- \kern -.04truein}
\def\narrowminussub{\kern -.02truein - \kern -.01truein}

\def\half{{1\over 2}}

\def\cl{\centerline}

\def\a{{\alpha}}
\def\g{{\gamma}}

\def\ph{{\phi}}

\def\l{{\lambda}}

\def\frac#1#2{{#1\over #2}}

\def\CN{{\cal N}}

\def\IZ{\relax\ifmmode\mathchoice
{\hbox{\cmss Z\kern-.4em Z}}{\hbox{\cmss Z\kern-.4em Z}}
{\lower.9pt\hbox{\cmsss Z\kern-.4em Z}}
{\lower1.2pt\hbox{\cmsss Z\kern-.4em Z}}\else{\cmss Z\kern-.4em
Z}\fi}
\def\IB{\relax{\rm I\kern-.18em B}}
\def\IC{{\relax\hbox{$\inbar\kern-.3em{\rm C}$}}}
\def\ID{\relax{\rm I\kern-.18em D}}
\def\IE{\relax{\rm I\kern-.18em E}}
\def\IF{\relax{\rm I\kern-.18em F}}
\def\IG{\relax\hbox{$\inbar\kern-.3em{\rm G}$}}
\def\IGa{\relax\hbox{${\rm I}\kern-.18em\Gamma$}}
\def\IH{\relax{\rm I\kern-.18em H}}
\def\II{\relax{\rm I\kern-.18em I}}
\def\IK{\relax{\rm I\kern-.18em K}}
\def\IP{\relax{\rm I\kern-.18em P}}

\def\p{\partial}

\font\cmss=cmss10 \font\cmsss=cmss10 at 7pt
\def\IR{\relax{\rm I\kern-.18em R}}

\def\e{{\epsilon_0}}

\def\at{ \tilde{\alpha}}

%

%
%
\def\eqnn#1{\xdef #1{(\secsym\the\meqno)}\writedef{#1\leftbracket#1}%
\global\advance\meqno by1\wrlabeL#1}
\def\eqna#1{\xdef #1##1{\hbox{$(\secsym\the\meqno##1)$}}
\writedef{#1\numbersign1\leftbracket#1{\numbersign1}}%
\global\advance\meqno by1\wrlabeL{#1$\{\}$}}
\def\eqn#1#2{\xdef #1{(\secsym\the\meqno)}\writedef{#1\leftbracket#1}%
\global\advance\meqno by1$$#2\eqno#1\eqlabeL#1$$}

\lref\rpol{J. Polchinski, ``TASI Lectures on D-Branes,''
hep-th/9611050\semi J. Polchinski, S. Chaudhuri and C. Johnson,
``Notes on D-Branes,'' hep-th/9602052. } 
\lref\rBFSS{T. Banks, W. Fischler, S. H. Shenker, and L. Susskind, ``M
Theory As A Matrix Model: A Conjecture,''  
hep-th/9610043, Phys. Rev. {\bf D55} (1997) 5112.}
\lref\rminwalla{S.Minwalla, ``Restrictions imposed by Superconformal Invariance on Qantum Field Theories''hep-th/9712074, Adv. Theor. Math. Phys., Vol 2 Issue 4.}
\lref\rstensor{A. Strominger, ``Open P-Branes,'' hep-th/9512059,
\pl{383}{1996}{44}.} 
\lref\rlu{J.X.Lu, ``ADM masses for black strings and p-branes,''
\pl{29}{1993}{313},hep-th/9304159.}
\lref\rkleba{S.Gubser, I.Klebanov and A.W.Peet, ``Entropy and temperature of 
Black 3-Branes'',hep-th/9602135 }
\lref\rklebb{I.Klebanov, ``World Volume Approach to absorption by Non Dilatonic Branes, hep-th/9702076 }
\lref\rklebc{S.Gubser, I.Klebanov,and A Teytlin ``String Theory and Classical Absorbtion
by Threebranes'',hep-th/9703040 }
\lref\rklebd{S.Gubser, I.Klebanov, ``Absorbtion by branes and Schwinger
terms in the World Volume Theory'',hep-th/9708005 }
\lref\rbrunner{I. Brunner and A. Karch, ``Matrix Description of
M-theory on $T^6$,'' hep-th/9707259.}
\lref\rmalda{J.Maldacena, ``The large N limit of Superconformal theories and
Supergravity,'' hep-th/9711020.}
\lref\rwit{E.Witten, ``Anti DeSitter Space and Holography,''
hep-th/9802150,.} 
\lref\rdpa{Dobrev and Petkova, ``Matrix Theory and U-Duality
in Seven Dimensions,'' hep-th/9702136 \pl{400}{1997}{260}.}
\lref\rfron{Sergio Ferrara, Christian Fronsdal , Alberto Zaffaroni,   ``On N=8
Supergravity on $AdS_5$ and N=4 SYM Theory'', hep-th/9802203.}
\lref\rroomana{A.Casher, F.Englert, H. Nicolai , M.Rooman, 
``The Mass Spectrum of Supergravity on the round seven sphere'',
\np{243}{1984}{173-188} }
\lref\rgunyadina{M.Gu\lref\rkallosh{P.Claus, R. Kallosh and A V Proeyen, 
``M-5 brane and superconformal(0,2) tensor multiplet in 6 dimensions ,''
hep-th/9711161}naydin and N.P. Warner, 
``Unitary Supermultiplets of OSp(8/4,R) ... Supergravity,'' 
\np{272}{1986}{99-124}}
\lref\rgunaydinb{M.Gunaydin and N. Marcus, 
``The Spectrum of the $S^5$ compactification of the Chira N=2, D=10... 
U(2,2/4),'' \cqg{2}{1985}{L11-L17}}
\lref\rgunaydinc{M.Gunaydin and P.  van Nieuwenhuizen, and N.P  Warner, 
``General construction of the unitary representaion of $AdS$ superalgebras
and the spectrum of the $S^7$ compactification of 11-dimensional Supergravity
'' \np{255}{1985}{63-92}}
\lref\rnieub{P.Nieuwenhuizen, cited as Ref 13 in \rgunaydinc}
\lref\rkallosh{P.Claus, R. Kallosh and A V Proeyen, 
``M-5 brane and superconformal(0,2) tensor multiplet in 6 dimensions ,''
hep-th/9711161}
\lref\rseiberg{N.Seiberg, 
`` Notes on Theories with 16 Supercharges''
hep-th/9705117}
\lref\rseibergb{O.Aharony, M Berkooz and N.Seiberg, 
`` Lightcone Description of (2,0) Superconformal Theories in Six Dimensions''
hep-th/9712117}
\lref\rpol{ S.Gubser, I. Klebanov and A Polyakov, 
`` Gauge theory Correlators from Non-Critical String Theory''
hep-th/9802109}
\lref\rooguri{ G Horowitz and H Ooguri,  
`` Spectrum of Large N Gauge Theory from Supergravity''
hep-th/9802109}
\lref\rstrominger{ J.Maldacena and A.Strominger, "$AdS_3$ Black Holes and a Stringy Exclusion Principle'', hep-th/9804085}

\Title{ \vbox{\baselineskip12pt\hbox{hep-th/9803053}
\hbox{PUPT-1773}}}
{\vbox{\centerline{Particles on $AdS_{4/7}$ and }
\centerline{Primary Operators on $M_{2/5}$ brane Worldvolumes}
}}

\smallskip
\centerline{Shiraz Minwalla\footnote{$^1$}{minwalla@princeton.edu}}
\medskip\centerline{\it The Department of Physics}
\centerline{\it Princeton University}
\centerline{\it Princeton, NJ 08540, USA}

\vskip 1in

\noindent 
I identify a correspondence between the various spherical harmonic
modes of massless 11 dimensional fields propagating 
on the $AdS_{4/7}$ in an $AdS_{4/7}\times S^{7/4}$ compactification of
M theory, and the corresponding operators, primary under the conformal group,
on the world volume of the $M_2, M_5$ branes. This is achieved by matching
representations of the superconformal algebra on the two sides of the 
correspondence. 

\vskip 0.1in
\Date{03/98}

\newsec{Introduction}

Recently Maldacena \rmalda\ observed that the duality between 
the `solution to low energy equation of motion' and gauge theory
(D brane) descriptions of solitons, implies a connection between
the world volume theory of some $p$ branes, and M (or string) theory
on $AdS_{p+2}\times S^{D-p-2}$. In particular he 
conjectured that the world volume theory
of the $M_{2/5}$ brane is dual to $M$ theory on $AdS_{4/7}\times S^{7/4}$.

Accepting the validity of his arguments, the next obvious 
task is to make the statement of the proposed duality more precise. 
This involves identifying a map between operators on the (superconformal)
world volume field theory and objects in the bulk theory, and 
stating the precise nature of this map (correlation functions on the world 
volume theory map onto what?)
These tasks have been partially accomplished for the special case of the 
$D_3$ brane in \rooguri , \rpol , \rwit , \rfron\ among others 

In this paper I identify a correspondence between the various spherical harmonic
modes of massless 11 dimensional fields propagating 
on the $AdS_{4/7}$ in an $AdS_{4/7}\times S^{7/4}$ compactification of
M theory, and the corresponding operators, primary under the conformal group,
on the world volume of the $M_2, M_5$ branes. This is achieved by matching
representations of the superconformal algebra on the two sides of the 
correspondence. 

The contents of the paper are as follows. In section two I state some 
properties of representations of superconformal algebras, and introduce
special representations that will be of interest to us. In section three 
I review papers on the particle  content of supergravity compactifications
on the background of interest, and note the results of
\rgunyadina , \rgunaydinc , grouping these particles into the superconformal
representations reviewed in section two. In section three I identify the 
superconformal primary operators on the world volume of the $M_2 / M_5$ 
branes that lead to same representations of the superconformal algebra as those
obtained on analyzing the representation content of particles
particles propagating on $AdS$ space in the corresponding supergravity  
theory. Descendents of 
these superconformal primary operators that are conformal primary operators
correspond to particles on the $AdS$  space. I give a reasonably 
easy to implement algorithm to determine an explicit expression for the 
world volume operator corresponding to any specific particle on the $AdS$
spaces.

\newsec{Special representations of the relevant superalgebras}

Unitary representations of the $D=3, \CN=8$ and $D=6, \CN=2$ superconformal 
algebras  have
been studied. Representations are infinite dimensional, and are completely
characterized by
a lowest weight multiplet of states. Lowest weights states appear in 
multiplets of $SO(3)\times SO(8)$ and $SO(6) \times SO(5)$ in the two cases
above respectively, and are labeled by the representations of  
these groups into which they fall, along
with a number $\e$, the scaling dimension of this representation. 
The condition of unitarity of representations
imposes an inequality, $\e \geq f(group\  \ rep)$, on the scaling dimension of 
allowed lowest weight states. The precise form of these inequalities has been
derived in \rminwalla . Representations at values of $\e$ saturating the
inequality above, and especially those representations that are at isolated
allowed values of $\e$,  are special short representations, containing 
fewer states than a generic representation. They are the 
analogues
of $BPS$ representations of Poincare supersymmetry.
Examples of such representations are 

a) Rep $Chiral_3(k)$:(D=3,\CN=8). $\e=\half k$;  $SO(3)=scalar$ ; $SO(8)=(k,0,0,0)$\foot{ I always 
specify $SO(k)$ weights in the GZ labeling system, in which the vector
is (1, 0, 0..), the spinor $(\half, \half, \half ...)$, etc}
in a convention 
(followed by \rgunyadina ) in which supersymmetry generators transform 
as $SO(8)$ chiral spinors.\foot{In \rminwalla\ scalar ultra short 
representations 
were derived to have scaling dimension $\e=h_1$ where $(h_1, h_2, h_3,h_4)$ 
denoted the $SO(8)$ representation. In that paper, however,
supersymmetry generators were taken to 
transform in the vector of $SO(8)$. To make contact with the results of that
paper one makes a triality transform , under which 
$(k,0,0,0) --> ({k \over 2}, {k \over 2}, {k \over 2},  {k \over 2})$.}

b) Rep $Chiral_6(k)$:(D=6,\CN=2) $\e=2k$, $SO(6)=scalar$ $SO(5)=(k,0)$.

$Chiral_3(k)$ and $Chiral_6(k)$ are ultrashort representations of the relevant superconformal 
algebras,
occurring at isolated values of allowed $\e$ as deduced in \rminwalla . These
particular representations will turn out to be of importance to us.

Gunaydin and collaborators have developed an oscillator technique to construct
representations of Lie Super Algebras. The representations above turn out to
be particularly easy to construct using this method. The construction has been
performed in \rgunyadina , \rgunaydinc , for $Chiral_3(k)$ and 
$Chiral_6(k)$ respectively. 

Any 
representation of the superconformal algebra may be decomposed into the sum
of finitely many representations of the conformal algebra\foot{In analogy with 
the fact that representations of supersymmetry consist of finitely many 
irreducible representations of the Poincare algebra, ie finitely many particles. BPS representations
are small because they contain few particles. Similarly ultrashort representations of the superconformal algebra are short because they contain few 
multiplets of the conformal algebra}. Gunaydin and collaborators have performed
this decomposition for the representations listed above. The results of this 
decomposition is listed in Table 1 in each of \rgunyadina , \rgunaydinc .

\newsec{ Superconformal representations and the compactification of 
d=11 SUGRA on $AdS_{4/7} \times S^{7/4}$.}

Consider $M$ theory compactified on $AdS_{4/7} \times S^{7/4}$. 
Purely classically, the massless states of M theory in 11
dimensions lead to infinite towers of `particles' in the 4/7 dimensional
$AdS$ spaces. Killing vectors of $AdS_{4/7}$ generate killing symmetries
of $SO(3,2) /SO(6,2)$. Equations of motion for all 4/7 dimensional particles 
commute with killing vectors; thus on shell modes of particles transform in 
representations (typically irreducible) of the $SO(3,2)/SO(6,2)$ killing 
symmetry group. Therefore on shell wave function modes of
elementary particles on $AdS_{4/7}$ appear in irreducible representations of 
$SO(3,2)/SO(6,2)$. 

The net symmetry supergroup of the whole 11 dimensional space
is $SO(3,2/8) /SO(6,2/2)$ \foot{The $SO(8)/SO(5)$ R symmetry is the 
killing symmetry of the sphere}, and so all modes on the compactified
11 dimensional spacetime must appear in multiplets the $(D=3,\CN=8)/(D=6,\CN=2)$ 
superconformal algebras, SO(3,2/8)/SO(6,2/2). Since any representation of a 
superconformal algebra may be decomposed into a sum of representations of the 
corresponding conformal algebra, and since each particle  on this space constitutes
 a representation
of the $d=3,6$ conformal algebra, particles must appear in families such the
on shell wave function of each family of particles constitutes an irreducible
representation of the d=3/6 superconformal algebras.

d=11 supergravity on $AdS_{4/7}\times S^{7/4}$ has been studied, and the 
resulting $d=4/7$ particle content extracted in \rroomana  , \rnieub . 
The particles thus obtained may be grouped into supermultiplets of the 
relevant superalgebras (this observation was made in \rgunyadina , \rgunaydinc ).
The representations of the superconformal algebras that these resulting 
particles lie in are all of, and only $Chiral_3(k)$/$Chiral_6(k)$, with $k\geq2$.

References \rgunyadina , \rgunaydinc\ explicitly decompose Reps $Chiral_3(k)$/$Chiral_6(k)$
into representations of the conformal group; to each such representation 
is a devoted a row in Table 1 of these references. 
Since particles on $AdS_{4/7}$ 
are irreducible representations of the relevant conformal algebra, there must be a one one 
correspondence between d=4/7 dimensional particles obtained from supergravity
and rows in Table 1 of \rgunyadina , \rgunaydinc\ for $k \geq 2$. 

The $d=4/7$ particles obtained by a compactification of $d=11$ SUGRA on 
$AdS_{4/7} \times S_{7/4}$ have been listed in \rroomana , \rnieub . One 
may try to establish a correspondence between the lists of particles in these
references, and the rows of Table 1 of  \rgunyadina , \rgunaydinc .
 
This correspondence has already been noted and made explicit in 
Table one of \rgunaydinc\ in the $d=7$ case. In the case $d=4$ the explicit
correspondence may be made by comparing Table 3 of \rroomana\ and Table 1
of \rgunyadina ;   specifically by comparing (column 1 , column 2) in
table 3 of \rroomana\ \foot{Note also formulae 3.10 and 3.11 in \rroomana\ .
These formulae list the mass of all 5 dimensional particles obtained on 
compactification as a function of the scaling dimension of the 
associated representation of the conformal group (what is referred to 
in that reference as the `Anti De Sitter Energy' of the representation).
$E_o$ in that paper is what I have called $\e$.} with (column 3, column 5) of \rgunyadina . When comparing
note that ($n$ of \rroomana\ ) = ($n$  of \rgunyadina ) minus 2. 

\newsec{{\bf The world volume theory of the $M_{2/5}$ brane}}

$M_{2/5}$ branes are governed by a world volume theories that are
superconformally invariant. Some information on 
 these theories may be found
in \rseiberg . 

The theory on the world volume of $m$ coincident
$M_{2}$ branes is the infrared limit of the world volume theory 
of $m$ coincident $D_2$ branes. The (0,2) theory of the $M$ 5-brane may 
also be connected to a gauge theory via flow in energy scale 
as follows. Consider an $M_5$ brane 
wrapped around the $M$ theory circle of radius $R$ in the decoupling limit
($M_{11}$ taken to $\infty$) at energy scale $E$.
At low energies
(or at fixed energy as $R \rightarrow 0$) the branes are governed by the 4 
dimensional world volume theory of $m$ $D_4$ branes at weak effective coupling
$=g_{ym}^2 E = RE$. At large $E$ (or fixed $E$ and large $R$) the world volume
theory is effectively that of the uncompactified (0,2) theory.  

Consider the world volume theory of a single $D_2$ brane. World volume
fields consist of 7 free real scalars, 8 free real spinors and a single
abelian gauge field. In 3 dimensions the abelian photon is 
dual to a compact scalar. In the infrared limit, the compactification
radius of the scalar goes to infinity, so in the infrared a
single $D_2$ brane is described by 8 real scalar and 8 real spinor degrees
of freedom. The scalars are taken to transform in the vector of $SO(8)$. The fermions
transform in the antichiral spinor of $SO(8)$. Supersymmetry generators
transform in the chiral spinor of $SO(8)$ \foot{ To keep conventions those
of \rgunyadina}. The Supersymmetry 
transformation laws are
\eqn\susya{\delta \l^{\at} ={i \over 2} \g^{\mu}\Gamma^a_{\a \at}\epsilon^{\a}
\p_{\mu}\phi^a} 
\eqn\susyb{\delta \phi^{a}={\overline \epsilon}^a\Gamma^a_{\a \at}\l^{\at} }  

The interacting theory of $m$ coincident 
$D_2$ brane may similarly be regarded as a theory of 8 matrix valued 
scalar fields, one of which is `compact' (and superpartners), 
even though in this case
the dualization of the photon into a scalar cannot be explicitly 
performed. 

In the interacting theory, the operators $Tr(X^{i_1}..X^{i_k})$, 
(where $i_1..i_k$ running 
from 1..7 are symmetrized and traceless) transform in the representation $(k,0,0..)$
of $SO(7)$, the internal symmetry group of $N=8$ SYM. The internal 
symmetry group of the theory is enhanced to $SO(8)$ in the infrared.
 Therefore the above multiplet of operators
must pair up with some other operators (consisting of symmetrized
traces of  $Xs$ and dualized photons) to form multiplets of $SO(8)$ 
in the infrared, transforming in $(k,0,0..0)$ of that group. 
I  denote these operators by $Tr(X^{j_1}..X^{j_k})$, 
where $j_1..j_k$ run from 1..8. I emphasize that by
$Tr(X^{j_1}..X^{j_k})$ I mean the operators to which these objects flow 
in the infrared - the actual infrared theory may have no convenient 
formulation in terms of $U(m)$ matrices of scalar fields. 

I now conjecture that, at the infrared fixed point, 
these operators are superconformal primary, and that their 
scaling dimension are given by $\e = {k\over 2}$. This implies
that these operators
head a superconformal multiplet of operators transforming in 
representation $Chiral_3(k)$ of the $D=3, \CN=8$ superconformal algebra. 
These operators and their descendents, therefore, correspond to the 
various modes of particles in the Kalutza Klein compactification of 
$M$ theory on $AdS_4\times S^7$.

$M$ theory on $AdS_4\times S^7$ also has multi particle supergravity states.
These states presumedly correspond, on the world volume of the $M_2$ brane, 
to products of the operators considered above. As $m$ is taken to infinity
however, the effective Planck Mass of the $M$ theory on $AdS_4\times S^7$ 
goes to 
infinity, and presumedly all nonsupergravity states attain infinite mass.
 This 
implies that the operators considered above along with their products 
exhaust the space of operators on the world volume theory at
$m = \infty$. 

Notice that the 
$D_2$ brane theory certainly possesses gauge invariant operators other than 
those considered above, for instance 
$Tr(X^iX^i)$, whose scaling dimension at very high energies (at which the gauge
theory is free) is 1. The $AdS$ correspondence predicts that the scaling 
dimension of this operator ( along with many others) runs off to infinity at low 
energies and $m=\infty$. This statement is plausible, because the effective
(t'Hooft) coupling of the theory scales like $m$, and so the large $m$ theory
is very strongly coupled. 
 
The story with the $M_5$ brane is similar.
The microscopic fields of the free theory of a single $M_5$ brane have  
5 scalars $\phi^a$ transforming as an $SO(5)$ vector, a single 
self
dual two form field $B_{\mu\nu}$, and 4 chiral spinors $\l$, 
transforming under 
$SO(5)$ as a chiral spinor. 
An explicit action for this theory may be found in section 4 of \rkallosh\ for 
instance. The supersymmetry transformation properties of the free theory are
listed in 
\rkallosh . The fields above transform under 
Rep $Chiral_6(1)$ of $SO(6,2/2)$.

The theory of $m$ coincident
$D_4$ branes possesses operators 
$Tr(X^{i_1}..X^{i_k})$ (with 
$i_1..i_k$ running from 1..5, and completely symmetrized and traceless) in the
representation $(k,0,0..0)$ of $SO(5)$. At very low energies these operators
have scaling dimension ${3\over 2 }k$. I conjecture that these operators
on the $D_4$ brane world volume are connected , via the renormalization 
group flow described at the beginning of this section, to primary operators
of scaling dimension $2k$ on the $(0,2)$ theory. These operators
head the infinite dimensional representation $Chiral_6(k)$ for all $k>1$,
of the $d=6, (0,2)$ superconformal algebra, and correspond to the 
supergravity particles propagating on $AdS^7\times S^4$ described in section 
3.   

The absence of other operators (save products of these) in the spectrum of the
infinite $m$ theory follows as for the 3 dimensional case. 

The spectrum of operators obtained in this theory  
agrees with the results 
of the DLCQ performed in \rseibergb\ , although the truncation of the spectrum of operators observed in \rseiberg\ 
does not emerge in any obvious fashion from supergravity, and is probably
related to a stringy exclusion principle \rstrominger\ . 

In the next section I will denote primary operators by a trace over 
$\ph $ fields. This notation refers to the primary 
operators on $M-branes$, connected to the 
specified trace operators in the relevant gauge theory
 by renormalization group flow. Conservatively this
description may be thought of as mere notation to keep track of
quantum numbers.

\newsec{{\bf Detailed Operator Particle Map}}

Having identified the primary operators under the superconformal group
corresponding to superconformal representations that occur in the 
$d=4/7$ particle spectrum of the previous section, it is
now easy to make a correspondence between specific particles on $AdS_{4/7}$
in the supergravity and specific descendents of one
of these operators. Since our primary operators are lowest weight
states in representations of the superconformal algebra that are
identical to those constructed in \rgunyadina , \rgunaydinb , 
one may obtain all conformally primary descendents of the superconformal
primary operators by a procedure identical to the one used in $\rgunyadina
\rgunaydinc$ in generating lowest weight states of the conformal algebra from
those of the superconformal algebra.   
Examining the procedure adopted in those papers leads to the following
algorithm

Consider the (superconformal) primary
operators that emerge from the trace over $k$ $\phi $ fields. Symmetrize
the $SO(8)/SO(5)$ indices of all these fields. This product is a primary 
operator under the superconformal algebra with as many descendent operators
that are  
primary operators of the conformal algebra as there are entries (for the 
relevant $k$) in table 1 of \rgunyadina, \rgunaydinc . Every entry 
(relevant to that $k$) in the table is associated with a specific conformal
primary operator. To construct the 
field corresponding to a given entry in that table, act on the product of
$\phi$s above with as many ($n$) $Q$ operators as there are boxes in either of 
the Young tableau appearing in column 2/1 of that table. Symmetrize the 
$SU(2)/SU(4)$ vector indices on the $Qs$ according to the first of the young tableau 
in column 2/1 of the table. Symmetrize the $SO(8)/SO(5)$ indices of the
$Qs$ according to the second of the young tableau  in column 2/1 of the 
table. \foot{The relative symmetrization of Lorentz and R symmetry indices
should be opposite - ie chosen such that the product of $Q s$ is completely
antisymmetric}. Multiply the $Q \  \ and \  \ \phi$
indices by $SO(8)/SO(5)$ Clebsch Gordan 
coefficients, and sum, so as to project onto the $SO(8)/SO(5)$ representation 
listed in column 4(or 5)/ 2(or 7) of that table. The operator thus obtained 
is a primary operator of the conformal group, with scaling dimension 
$\e ={n+m \over 2}\  \  / \  \  2n+{m \over 2} $, $SO(3)/SO(6)$ transformations as given by the first 
Young Tableau in column 2/1 of the table, and appearing in $SO(8)/SO(5)$ 
multiplets listed in columns [4 and 5] /[2 and 7] of the table. 

I elaborate on this process for some simple examples to clarify procedure.
Consider the $AdS_4$ graviton. This corresponds to the only spin 
2 particle in Table 1 of \rgunyadina . Making reference to the mass formula
in \rroomana , we note that the massless graviton corresponds to $k=2$ in
\rgunyadina . Using Table 1 in that reference, we thus see that the 
graviton is obtained by acting the product of 4 $Q$ operators on the trace
of two $\phi  $ fields. According to the table and the algorithm above, the
4 $SU(2)$ spinor indices on these $Qs$ must be completely symmetrized.  That
makes sense as it yields a spin 2 particle. According to the table, the
4 $SO(8)$ chiral spinor indices on the $Qs$ must be completely antisymmetrized.
This process leads to two $SO(8)$ representations,  the symmetric tensor, 
and the anti selfdual 4 form. Since we are (according to the table) supposed
to couple this representation to the symmetric tensor (from the $\phi s$ ) to
form a scalar, we select out the symmetric tensor, and contract with the
symmetric tensor indices on $\phi s$ to get an $SO(8)$ scalar. Following 
through this procedure, and using the explicit expressions above for the
action of the supersymmetry generators on fields yields an explicit 
expression for the world volume field corresponding to the $d=4$ graviton.

As another example consider the $AdS_7$ massless vector particle. It 
corresponds to the entry in row 4 of Table 1 in \rgunaydinc . It occurs 
at $p=2$ in that table (according to the mass formula in the table). 
According to the table, it is formed by acting on the trace on a product 
of 2 $\phi s$ by 2 $Q$ s. Antisymmetrize the chiral $SO(6)$ spinor
indices on these $Qs$ obtaining an $SO(6)$ vector. Symmetrizes the 
$SO(5)$ spinor indices  on these $Qs$ obtaining an antisymmetric tensor.
Finally Clebsch Gordan couple the antisymmetric tensor with the symmetric 
tensor (from  the $\phi s$ ) to get an antisymmetric tensor (according to
the table).

As a last example consider the $AdS_7$ graviton. It corresponds to row 6
in Table 1 of \rgunaydinc , $p=2$. One obtains it by acting on the 
trace of 2 $\phi s$ with 4 $Q s$. The $Q$ $SO(6)$ indices are pair wise 
anti symmetrized to form 2 sets of $SO(6)$ vector indices, and these
vector indices are mutually symmetrized. The $Q$ $SO(5)$ spinor indices
are also pairwise antisymmetrized to form $SO(5)$ vectors (discard the scalar
component), but the pairs are chosen incommensurately with  the first set of
pairs. The resulting 2 $SO(5)$ vector indices are then symmetrized to form
a symmetric tensor,  which is then contracted with the symmetric tensor indices
of the $\phi s$ to yield an $SO(5)$ scalar

\newsec{Conclusion}

In this paper I have mapped chiral superconformal primary operators of 
world volume theories onto groups of particles appearing in the 
corresponding Maldacena dual SUGRA on $AdS_{4/7} \times S^{7}$. In the 
process, I have 
computed the spectrum of chiral operators on the 
worldvolume of the $M_{2/5}$ brane. The spectrum so obtained agrees with
the analysis of $\rseibergb$ for the $M_5$ brane, and so may be regarded
as a test of Maldacena's conjecture.

\cl{\bf Acknowledgments}

I would like to thank M. Gunaydin for a communication bringing his interesting
work on oscillator constructions of supergroups to my attention. 
I would like to acknowledge useful discussions with 
I.Klebanov , M.Krogh, S.Ramgoolan, S.Lee
and A.Mikhailov. I would like especially to thank N. Seiberg for advice,
encouragement and numerous helpful comments.

\listrefs
\end